\documentclass{aa}  
\usepackage{graphicx}
\usepackage{txfonts}

\usepackage{natbib}
\bibpunct{(}{)}{;}{a}{}{,} 

\begin{document}
   \title{The distance to the cool T9 brown dwarf \object{ULAS~J003402.77-005206.7}}

   \author{R. L. Smart\inst{1}
          \and
          H. R. A. Jones\inst{2}
          \and
          M. G. Lattanzi\inst{1}
          \and          	
          S. K. Leggett\inst{3}
          \and
          S. J. Warren\inst{4}
          \and
          A. J. Adamson\inst{5}
          \and
          B. Burningham\inst{2}
          \and
          M. Casali\inst{6}
          \and
          D. W. Evans\inst{7}
          \and
          M. J. Irwin\inst{7}
          \and
          D. Pinfield\inst{2}
}

   \offprints{R.L. Smart}

   \institute{
   Istituto Nazionale di Astrofisica, Osservat\'orio Astronomico di Torino,
              Strada Osservat\'orio 20, 10025 Pino Torinese, Italy
        \and
   Centre for Astrophysics Research, Science and Technology Research Institute, University of Hertfordshire, Hatfield AL109AB, UK
   \and
   Gemini Observatory, 670 N. A'ohoku Place, Hilo, HI 96720, USA
   \and
   Imperial College London, Blackett Laboratory, Prince Consort Road, London
   SW7 2AZ, UK
   \and
   Joint Astronomy Centre, 660 North A'ohoku Place, Hilo, HI 96720, USA
   \and
   ESO, Karl-Schwarzschild-Str. 2, D-85748 Garching bei Munchen, Germany
   \and
   Institute of Astronomy, Madingley Road, Cambridge CB3 0HA, UK
}

   \date{Received 12/11/2009, accepted 16/12/2009}
 
    \abstract
     {}
     {We demonstrate the feasibility of determining parallaxes { for
         nearby objects} with the Wide Field Camera on the United Kingdom
       Infrared Telescope (UKIRT) using the UKIRT Infrared Deep Sky Survey as
       a first epoch. We determine physical parameters for
       ULAS~J003402.77-005206.7, one of the coolest brown dwarfs currently
       known, using atmospheric and evolutionary models with the distance
       found here.}
     {Observations over the period 10/2005 to 07/2009 were pipeline processed
       at the Cambridge Astronomical Survey Unit and combined to produce a
       parallax and proper motion using standard procedures. }
     {We determined $\pi = 79.6 \pm 3.8$ mas, $\mu_{\alpha} = -20.0 \pm
       3.7$ mas/yr and $\mu_{\delta} = -363.8 \pm 4.3$ mas/yr for ULAS
       J003402.77-005206.7.  }
     {We have made a direct parallax determination for one of the coolest
       objects outside of the solar system. The distance is consistent with a
       relatively young, 1 -- 2 Gyr, low mass, 13 -- 20 M$_{J}$, cool,
       550-600K, brown dwarf.  {We present a measurement of the radial
         velocity that is consistent with an age between 0.5 and 4.0 Gyr. } }

\titlerunning{The distance to ULAS~0034}

   \maketitle
%

\section{Introduction}

The prototype T dwarf was discovered in 1995 as a companion to the nearby M
dwarf star Gl229 (Nakajima et al. 1995), and today we know of 155 T dwarfs
(www.dwarfarchives.org as of 10/2009). The majority of these are a result of
the near-infrared 2 Micron All Sky Survey \citep[2MASS]{2006AJ....131.1163S}
and the deep optical Sloan Digital Sky Survey
\citep[SDSS]{2000AJ....120.1579Y}.  Once discovered, significant efforts were
undertaken to determine their distances \citep{2004AJ....127.2948V,
  2003AJ....126..975T, 2002AJ....124.1170D} to map out the lower end of the
Hertzsprung-Russell (H-R) diagram and to constrain models. However, in these
surveys only a handful of objects have spectral types T7 or later and more
examples of intrinsically fainter and cooler objects are crucial to understand
both such objects and, in turn, exoplanets of similar masses.

In 2005 the UKIRT Infrared Deep Sky Survey
\citep[hereafter UKIDSS]{2007MNRAS.379.1599L} began 
with the Wide Field Camera \citep[WFCAM]{2007AA...467..777C} large-field
infrared camera. This survey is going about{ three} magnitudes fainter than the
2MASS survey and has already revealed a number of extremely faint and cool T
dwarfs \citep{2009MNRAS.397..258L, 2008MNRAS.391..320B, 2008MNRAS.390..304P,
  2007MNRAS.379.1423L}. As part of the followup for the UKIDSS surveys we have
started a program to determine parallaxes for the coolest objects being
discovered.

The first and most obvious application of measured distances is to derive a
luminosity which is used to populate the H-R diagram and constrain 
models. Distances are also used to identify unresolved companions via
over-luminosities, and to provide space motions which provide indications of
an object's origin and age. 

There is a large degree of uncertainty in current model analyses of the
late-type T dwarfs, due to known inadequacies in the atmospheric models
(e.g. Leggett et al. 2009). However, evolutionary models of cooling brown
dwarfs are well understood, and show that all brown dwarfs older than about
200 Myr have a radius within about 20\% of Jupiter's
\citep[e.g.][]{2001RvMP...73..719B}. Hence there is a tight relationship between
luminosity and effective temperature ($T_{\rm eff}$) for brown dwarfs (via the
Stefan-Boltzman law). Parallax determinations combined with flux-calibrated
observed spectra provide luminosities, and hence the best measurements of
$T_{\rm eff}$ for brown dwarfs currently available.  Once $T_{\rm eff}$ is
known, gravity and metallicity can be constrained by comparing synthetic
spectra to the observations
\citep[e.g.][]{2007ApJ...656.1136S}. \citet{2009MNRAS.395.1237B} use luminosity
to determine $T_{\rm eff}$ for the T8.5 Wolf 940B, and show that an analysis
using synthetic spectra alone would have overestimated the temperature by
$\sim$100~K, for this 600~K object.

The latest spectral type currently defined is T9 and three such objects are
currently known: ULAS {J003402.77-005206.7} \citep[hereafter ULAS
0034]{2007MNRAS.381.1400W}, CFBDS J005910.90-011401.3
\citep{2008AA...482..961D} and ULAS~J133553.45 +113005.2
\citep{2008MNRAS.391..320B}. They have effective temperatures in the range 500
to 700~K, 
and masses of 5 to 50 M$_J$ corresponding to ages of 0.1 to 10 Gyr
(e.g. Leggett et al. 2009).  Here we present the first results of the UKIRT
parallax program for ULAS~0034.  In Sect. 2 we describe the observations and
procedures for the program, in Sect. 3 we provide the results for ULAS~0034
and in Sect. 4 we find the most consistent set of physical parameters for
ULAS~0034 by combining available observations and models.

\section{Observations and Reduction Procedures }

\subsection{Imaging}
{The imaging }observations were all made on the UKIRT 3.8~m telescope
using the WFCAM imager. This is the instrument being used to carry out
the UKIDSS surveys, and the calibrated images were taken directly from
the pipeline for that survey{ \citep{2004SPIE.5493..411I}.}  All
observations are carried out in queue override mode, allowing us to be
very flexible in the scheduling, maximising the parallax factor and
observing close to meridian passage. The observational sequence we
adopt is very similar to the $J$-band imaging sequence in the Large
Area Survey (LAS) component of UKIDSS
\citep[e.g.][]{2007MNRAS.375..213W}.  A longer exposure time was used
to provide an increased signal-to-noise for the targets, which are in
general at the faint limit of the LAS. Specifically, we make
observations in a 5 jitter (dithered) 3.2'' cross pattern, and at each
jitter position we make 4 exposures in 2$\times$ 2 micro-stepped
positions of 1.5 pixels, where each exposure consists of 2 co-added
10~second images. The total exposure time is therefore $5\times 4
\times 2 \times 10 = 400$~seconds. In average conditions this provides
a signal-to-noise of 100 at $J=18$.

ULAS~0034 was discovered in a sequence of images taken on 2005/10/04. We use
the same pointing, filter and detector setup as the $J$ image from that night
in our parallax sequence. This allows us to use the discovery image as the
first epoch in the parallax determination, and we reduce the time required for
that determination by about one year (the time between the discovery
observation and the recognition that an object warrants the investment
required to determine a parallax).  Nominally, to distinguish between
parallactic and proper motion we aim to have observations that cover at least
three years, hence the saving of one year is significant.

\begin{figure}
  \centering
  \includegraphics[width=9cm]{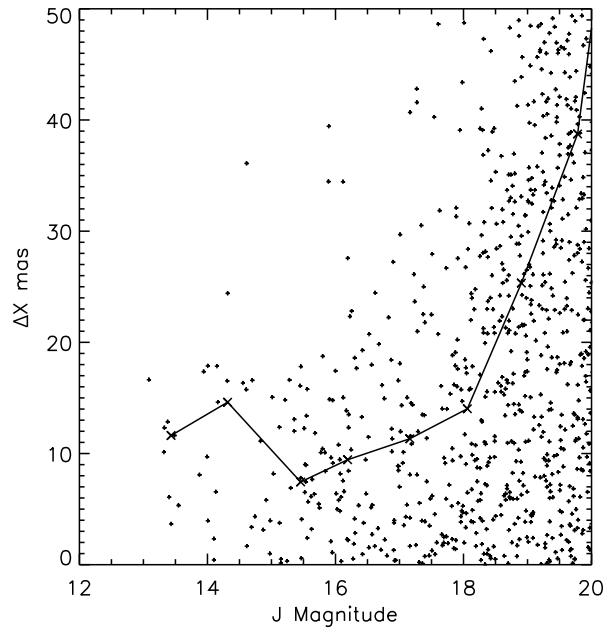} 
  \caption{Object position difference between two observations of the same
    cluster made on 2005/06/20 and 2005/09/04 as a function of $J$
    magnitude. Plus signs are individual objects, crosses are median values.}
  \label{wtest_for_paper1.eps}%
\end{figure}

Parallax observations require precise astrometry. Unfortunately WFCAM has a
large field-of-view with significant radial distortion, and as all four
detectors are offset from the optical axis they are subject to large
astrometric distortions. To minimize the differential astrometric distortion
between the discovery and subsequent observations, the original pointing
orientation is always used unless the object is close to the edge of a
detector, in which case we move the object towards the centre. We work under
the assumption that the astrometric distortion does not change for the
duration of the observational program. This assumption forms the basis of all
small-field, high-precision, ground-based parallax programs, even in
telescopes where the focal plane is considered astrometrically flat, and is
required to be true for WFCAM because of the known large astrometric
distortions \citep{2004SPIE.5493..411I}.

During our parallax campaign WFCAM, a forward Cassegrain focus instrument, was
regularly unmounted for significant periods so that the Cassegrain instruments
could be used.  To check that unmounting and remounting does not lead to
significant changes in the astrometric map, a comparison of the observations
of the same cluster field for the nights 2005/06/20 and 2005/09/04 were made,
which straddles a period where WFCAM was unmounted and remounted. In
Fig. \ref{wtest_for_paper1.eps} we plot the differences in local x
coordinates, as derived from the Cambridge Astronomical Survey Unit (CASU)
pipeline discussed below. The first night was adopted as the reference system
and the second night transformed using a six-constant model (translation,
scale and rotation) derived from common objects. Assuming the nights were of
equal quality the per observation error will be approximately equal to the
combined error divided by $\sqrt{2}$. From this comparison we estimate that
the per observation precision for objects to $J=18$ is around 10~mas in both
coordinates including any contribution from the unmounting and remounting of
WFCAM. An examination of binned residuals as a function of detector position
did not reveal any systematic patterns.

\begin{figure}
  \centering
  \includegraphics[width=9cm]{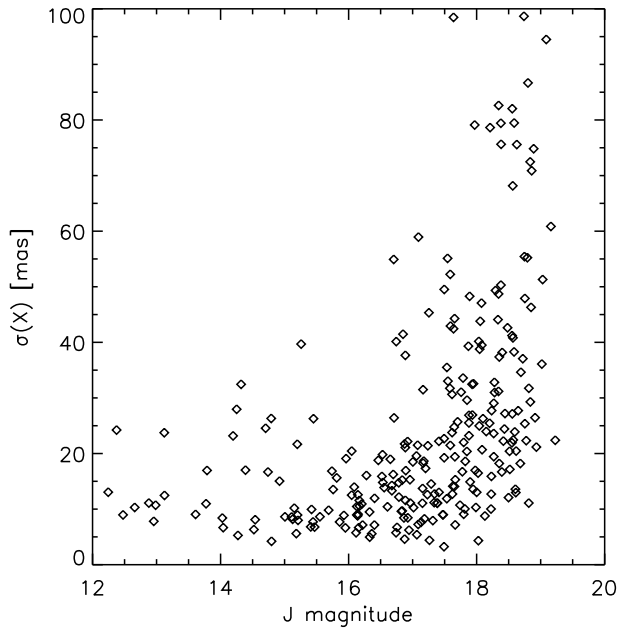} 
  \caption{Root-mean-square of the x coordinates for stars in 
    the first 8 images of the ULAS~0034 sequence which spans 2.6 years, as a
    function of $J$ magnitude. The centroids were all derived using the CASU
    pipeline.} 
  \label{f4800792rwf_fortest0cx.eps}%
\end{figure}

The observations made with WFCAM are processed and reduced using a
dedicated pipeline being run by the CASU. Since the centroid
of the objects is our fundamental observation, we have extensively tested the
results of this pipeline and we now describe these tests.

First we have taken the images as they come out of the CASU pipeline
(i.e. dark-subtracted, flat-fielded, systematic noise removed, sky-corrected,
and finally registered and stacked) and found centroids two ways: (1) fitting
a two-dimensional gaussian to the object's point source function (PSF) and (2)
fitting a one-dimensional gaussian to the object's marginal distributions
above the sky background. The two-dimensional gaussian fit is what is used in
the Torino Observatory Parallax Program \citep[hereafter TOPP]{SMA99A} and the
marginal distribution fit is what is suggested by Stone (1989)
\nocite{1989AJ.....97.1227S} in the presence of a high background. We use
these centroids and those from the CASU pipeline to compare eight ULAS~0034
observations over the period 2005 to 2008, after a six-constant transform
(translation, scale and rotation) to a common system.  In Fig.
\ref{f4800792rwf_fortest0cx.eps} we plot the root-mean-square (rms) of the x
coordinate from the CASU pipeline of common stars. The median rms is 18 mas
and using the other centroiding methods the rms was identical. In the y
coordinate the median rms was slightly larger, 20 mas, but the difference
between the three centroiding methods was also negligible.

We have also tried recombining the original micro-stepped images using a
drizzle \citep{2002PASP..114..144F} routine rather than the CASU standard
routine. In the standard routine the 4 micro-stepped images are interleaved,
i.e. combined assuming a higher spatial resolution, to produce a combined
image with pixel size 0.2'' (half the 0.4'' physical pixel size). This
produces a spiky PSF as the seeing changes between the four images. The
alternative approach is to combine the counts in pixels from the four images
in an underlying higher resolution grid with appropriate weighting
(i.e. drizzling). This gives a much smoother PSF and retains the
signal-to-noise, while losing some of the resolution as adjacent pixels are
correlated. Using this process caused the ULAS~0034 median rms to increase
slightly from 18 to 20 mas.
 
As a result of these tests we have decided to use positions coming directly
out of the CASU pipeline, and work under the assumption that our median error
is around 18 mas.  We assume the larger error found in the ULAS~0034 sequence
compared to the two-epoch cluster test is due to the fact that we are comparing a wider
range of observing conditions over a longer time span. We note that the
centroiding errors provided by the CASU pipeline are good indications of the
errors found in our test.

Once the (x,y) coordinates have been determined, the parallax and proper motion
of ULAS~0034 are determined using the methods adopted in the TOPP
\citep{Sma03a, 2007AA...464..787S}. We limit the reference objects used to
only those within 5 arcminutes of the target, as these are sufficient for a
transformation, and limiting the area of the detector being modeled also limits
possible differential astrometric distortion.  We have made two significant
changes to the procedures used in the TOPP:\\
(1) As we are observing in the $J$-band where the atmospheric refraction is small
 we expect the differential reddening correction (DCR), to be
negligible. This is in agreement with \citet{2003AJ....126..975T} and hence we
do not apply any correction. Once we have data on more targets we
will estimate the DCR and quantify the effect this could have on
the target parallaxes.\\
(2) Since we have the SDSS colours in this region we use the method of
\citet{2008ApJ...684..287I} to determine photometric parallaxes for the
anonymous reference stars and from those we calculate the correction from
relative to absolute parallax. This correction is typically less than 2
milliarcseconds.

{ In the field there are also significant numbers of galaxies which
  can be used to correct from a relative to an absolute system. For the proper
  motions we found the mean of the galaxies was less than the error of that
  mean and hence, within their error, these proper motions are on an absolute
  system. For the parallaxes we would need to use the parameters of the
  galaxies as constraints \citep{Eic97A}, future work will investigate this
  possibility.  }

\subsection{Spectroscopy}

{
We have also obtained new near-infrared spectra of ULAS 0034 as well as of the
template T8 dwarf 2MASS\,J0415 \citep{2002ApJ...564..421B}, for comparison. By
spectral type T8 the CH$_4$ and H$_2$O molecular bands that define the T
sequence are practically saturated. Therefore, despite the low effective
temperature of ULAS 0034, $550-600$K (see next section), relative to 750K for
2MASS\,J0415 \citep{2007ApJ...656.1136S}, at low resolution the near-infrared
spectral differences are reasonably subtle. For this reason we obtained
medium-resolution spectra in the $Y$ and $J$ bands to search for any spectral
differences identifiable at the higher resolution. One possibility is the
appearance of absorption lines due to NH$_3$ \citep{2007ApJ...667..537L}.

The observations were made with the ISAAC instrument on the ESO VLT in
service mode, during several different nights within the period July to
September 2008. We used a $1\arcsec$ slit and the medium-resolution (MR)
grating, providing wavelength coverage of $0.999-1.046\mu$m in the $Y$
band, and $1.199-1.258\mu$m in the $J$ band. Total integration times for
ULAS 0034 were 160\,min in the $Y$ band and 120\,min in the $J$ band, and
for the much brighter source 2MASS\,J0415 were 40\,min in the $Y$ band
and also 40\,min in the $J$ band. We previously obtained a spectrum with
ISAAC in the H band, of similar resolution, and the data reduction is
described in Warren et al (2007b). We followed the same procedures for
reducing the $Y$ and $J$ spectra here.

When smoothed to the lower resolution of the original GNIRS spectrum
(from Warren et al., 2007b) we find the new spectra show good
agreement, confirming in particular the additional absorption for ULAS
0034 compared to 2MASS\,J0415, in the blue wing of the $J$
band. Nevertheless a search for individual absorption lines failed to
reveal any new spectral features.

The new $J$-band spectrum of ULAS 0034 is of substantially higher S/N
than the $Y$-band spectrum, and we use it here to measure the
heliocentric radial velocity. The average seeing, $0.85\arcsec$, was a
little less than the slit width, and determines the wavelength
resolution of the spectrum, $3.5\times10^{-4}\mu$m FWHM. The use of a
slit width larger than the seeing is not ideal for the measurement of
a radial velocity (which was not the primary purpose of the
observations), since any mis-centering of the source in the slit will
introduce a systematic error. We quantify this error in the next
section, when discussing the radial velocity.

}

\section{Results}

\begin{figure}
  \centering
  \includegraphics[width=9cm]{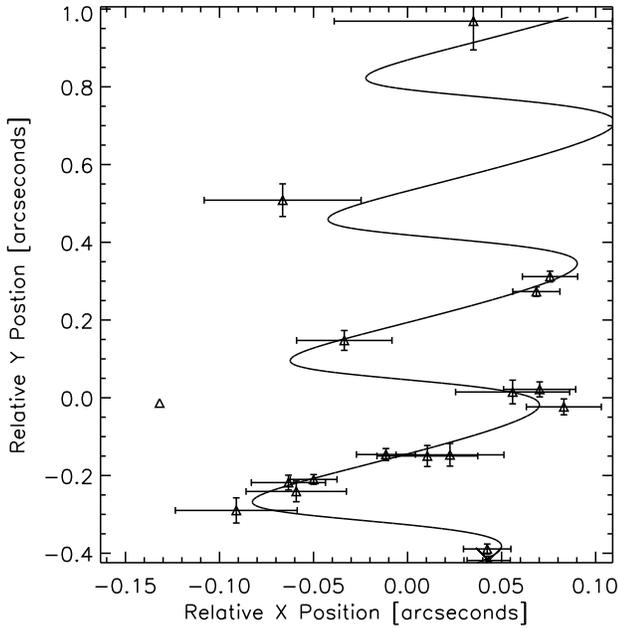}
  \caption{Observations and solution for ULAS~0034. The point at (0.03,
    1.0) is the discovery observation on 2005/10/04 which had slightly lower
    signal-to-noise than subsequent observations, hence the larger error. The
    observation without error bars was made in{ twilight and in poor
    seeing, and is of low quality,} and was rejected from the solution
    by three-$\sigma$ clipping.} 
  \label{T00345_par.ps_page_22}%
\end{figure}

In {Table} \ref{results} we report the astrometric parameters derived for ULAS
0034. In Fig. \ref{T00345_par.ps_page_22} we plot the
observations and solution derived. 

\begin{table}[ht]
\caption{Solution parameters for ULAS~0034 and corresponding absolute magnitudes}
\begin{tabular}{ll}
\hline  \\
 $\alpha, \delta$ ~~ (J2000.S)          & 00:34:02.70, -00:52:07.7 \\ 
 Base epoch (yr)                        &       2008.1312 \\ 
 Absolute parallax (mas)                &  79.6 $\pm$ 3.8  \\ 
 $\mu_{\alpha}$ (mas/yr)                &  -20.0 $\pm$  3.7 \\ 
 $\mu_{\delta}$  (mas/yr)                &  -363.8 $\pm$  4.3  \\ 
 Relative to absolute correction (mas)  & 1.33 \\ 
 Duration of observations, (yrs)        & 3.81 \\ 
 observations, reference stars & 17, 113 \\ 
 Absolute Magnitude in J$^1$          & 17.65 $\pm$  0.11 \\
 Absolute Magnitude in K$^1$          & 17.98 $\pm$  0.12 \\
\hline 
\label{results}
\end{tabular}

\small{$^1$ Absolute magnitudes based on apparent magnitudes published in 
Warren et al. (2007b) and including both parallax and apparent magnitude
errors.}
\end{table} 

{
We measured the heliocentric radial velocity of ULAS~0034 relative to
2MASS\,J0415 by cross-correlating the two J-band spectra, finding a
value $-43.6 \pm 5.4$ km/s, i.e. ULAS~0034 is blue-shifted relative to
2MASS\,J0415.  The quoted uncertainty accounts for the S/N of the two
spectra, but not for the systematic error, due to mis-centering of one
or other object in the slit. We estimate, conservatively, that the
accuracy of centering the object in the slit was as poor as
$0.2\arcsec$. Then the systematic velocity error, for a single
measurement, would be $\sim$10km/s. We computed this value by
establishing the shift of the centroid of the light captured by the
slit, given the measured seeing, and assuming a Gaussian profile. This
error dominates the error budget for the radial velocity, and it would
be possible to obtain a more accurate velocity with a different
instrumental setup. Combining in quadrature the systematic error for
two measurements with the random error, we obtain a total uncertainty
of 15 km/s. Taking the mean of the (much more accurate) heliocentric
velocities for 2MASS\,J0415 given in \citet{2007ApJ...666.1205Z}, we
}conclude that ULAS~0034 has a radial velocity of $+7 \pm 15$
km/s. Combining this with the measured proper motions, parallax and
the velocity of the Sun from \citet{1998MNRAS.298..387D} we find a
space motion, in km/s, of U=$23 \pm 3$ (radially inwards), V= $-8 \pm
6$ (in the direction of Galactic rotation), and W=$-9 \pm 13$
(vertically upwards).

Finally, we compare apparent $J$ magnitudes for our 3 year parallax sequence
and find an internal standard deviation of 0.02 magnitudes and a systematic
variation of less than 0.01. Hence ULAS~0034 does not appear to be
photometrically variable in the near-infrared.

\section{Discussion}

In the ULAS 0034 discovery paper, \citet{2007MNRAS.381.1400W}, they use near-infrared
spectroscopy and mid-infrared photometry, with model atmospheres, to derive
parameters of $600 \leq T_{\rm eff}$~K$ \leq 700$ and $4.5 <$ log $g < 5.1$
for ULAS~0034, corresponding to a mass range of 15 to 36 M$_J$ and an age
range of 0.5 to 8 Gyr.  The inferred photometric distance is 14 -- 22 pc.
\citet{2008AA...482..961D} use near-infrared spectra alone, with models, to
estimate $T_{\rm eff} \approx 670$~K and log $g \approx 4.6$ for ULAS~0034,
consistent with the \citet{2007MNRAS.381.1400W} analysis. \citet[hereafter
LCS09]{2009ApJ...695.1517L} used a model comparison to a more complete
spectral energy distribution which included mid-infrared spectroscopy to
constrain the physical properties of ULAS~0034 further.  They find that $550
\leq T_{\rm eff}$~K$ \leq 600$, $4.0 <$ log $g < 4.5$ and $0.0 <$ [m/H] $<
+0.3$. These values implied a distance of 13 -- 16 pc, a mass of 5 -- 20 M$_J$
and an age of 0.1 -- 2 Gyr. The parallax determined here allows us to further
refine these results.
  
The measured distance of 12.6 $\pm$ 0.6, rules out the
low gravity (log $g = 4.0$), very low mass (5 -- 8 M$_J$) and very young age
(0.1 -- 0.2 Gyr) LCS09 fits to ULAS~0034.  We can check that the distance
derived here is consistent with the other solutions by deriving the bolometric
luminosity and comparing this to the family of {$T_{\rm eff}$, log $g$}
solutions calculated by structural models for a given luminosity (e.g. Saumon
\& Marley 2008, their Fig. 4)\nocite{2008ApJ...689.1327S}.

We derived the bolometric flux by interpolating between the flux calibrated
0.8 -- 2.3 $\mu$m GNIRS and 8 -- 15 $\mu$m IRS spectra presented in LCS09
using {$T_{\rm eff}=550$~K, [m/H] $= +0.3$} and {$T_{\rm eff} = 600$~K, [m/H]
  $= 0.0$} log $g = 4.5$ synthetic spectra.  The 2.3 -- 8 $\mu$m region
contributes 33 -- 43\% of the total flux. The contribution beyond 15 $\mu$m
was determined assuming a Rayleigh-Jeans tail to the distribution and $T_{\rm
  eff} = 550$~K or 600~K; this region contributes $\sim$8\% of the total flux.
The model interpolation and the Rayleigh-Jeans extrapolation were scaled by
adopting the distance measured in this work, and the structural model radius
of 0.11 R$_{\odot}$ as per LCS09. We find a luminosity of log L/L$_{\odot} =
-5.96 \pm 0.06$, including the 4.8\% uncertainty in distance and an estimated
total flux uncertainty of 13\%. The flux uncertainty is dominated by the
estimated 13\% uncertainty due to the model interpolation section; other
sources of error are the uncertainty in the flux calibration of the observed
spectrum (3\%) and the assumption of a Rayleigh-Jeans tail (2\%). The
luminosity is consistent with the $T_{\rm eff} = 550$ -- 600~K, log $g = 4.5$
LCS09 solution for ULAS~0034, implying an age of 1 -- 2 Gyr and a mass of 13
-- 20 M$_J$. Note that the measured parallax rules out the possibility that
ULAS~0034 is a 550 -- 600~K binary system as the distance would have to be
larger than any of the LCS09 solutions. Finally we note the velocity and age
are consistent with the younger populations ($<$ 4 Gyr) in Figs. 5 \& 6 of
\citet{2009A&A...501..941H} but outside the very young object ($<$ 0.5 Gyr)
velocity box of Fig. 6 in \citet{2004ARA&A..42..685Z}. 
{
\begin{table}[ht]
\begin{center}
\caption{~~~~Summary of physical parameters of ULAS~0034}
\begin{tabular}{ll}
\hline  
 Distance (pc)     &    12.6 $\pm$ 0.6    \\
 U,V,W    (km/s)      &    $23 \pm 3$, $-8 \pm 6$, $-9 \pm 13$ \\
 T$_{\rm eff}$ (K)  &   550 -- 600 \\
 log $g$       &  4.5 \\
 Mass (M$_J$  )         & 13 -- 20  \\
 Age  (Gyr)         & 1 -- 2 \\
\hline 
\label{prop}
\end{tabular}
\end{center}
\end{table} 
}

The current UKIDSS followup program has 11 targets which will result in
equally precise parallaxes by the end of 2009. Following the success for ULAS
0034 we recently added 20 other faint T dwarfs which will have enough
observational coverage for a parallax determination by the end of 2011.

\section{Acknowledgments}
RS and HRAJ would like to acknowledge the support of Royal Society
International Joint Project 2007/R3. This research has benefitted from the M,
L, and T dwarf compendium housed at dwarfArchives.org and maintained by Chris
Gelino, Davy Kirkpatrick, and Adam Burgasser.  We would like to express our
thanks to Luca Rizzi for scheduling help and all the observers at UKIRT who
have carried out the observations used here.  The United Kingdom Infrared
Telescope is operated by the Joint Astronomy Centre on behalf of the Science
and Technology Facilities Council of the U.K. SKL's research is supported by
the Gemini Observatory, which is operated by the Association of Universities
for Research in Astronomy, Inc., on behalf of the international Gemini
partnership of Argentina, Australia, Brazil, Canada, Chile, the United
Kingdom, and the United States of America.{ The ESO ISAAC spectra
  described here were obtained under programme 381.C-0036(A).}

\bibliographystyle{aa} 
\bibliography{local}

\end{document}